# Targeted Backdoor Attacks on Deep Learning Systems Using Data Poisoning

Xinyun Chen  Chang Liu  Bo Li  Kimberly Lu  Dawn Song
UC Berkeley   UC Berkeley   UC Berkeley   UC Berkeley   UC Berkeley

*Abstract*—Deep learning models have achieved high performance on many tasks, and thus have been applied to many security-critical scenarios. For example, deep learning-based face recognition systems have been used to authenticate users to access many security-sensitive applications like payment apps. Such usages of deep learning systems provide the adversaries with sufficient incentives to perform attacks against these systems for their adversarial purposes.

In this work, we consider a new type of attacks, called *backdoor attacks*, where the attacker's goal is to create a *backdoor* into a learning-based authentication system, so that he can easily circumvent the system by leveraging the backdoor. Specifically, the adversary aims at creating backdoor instances, so that the victim learning system will be misled to classify the backdoor instances as a target label specified by the adversary.

In particular, we study *backdoor poisoning attacks*, which achieve backdoor attacks using poisoning strategies. That is, the attacker injects *poisoning samples* into the training set to achieve his adversarial goal. Different from all existing work, our studied poisoning strategies can apply under a very weak threat model: (1) the adversary has no knowledge of the model and the training set used by the victim system; (2) the attacker is allowed to inject only a small amount of poisoning samples; (3) the backdoor key is hard to notice even by human beings to achieve stealthiness. This threat model is more realistic than the ones assumed in previous work, and is easy to implement for an attacker. Satisfying all these constraints is challenging, and our work is the first one to show the feasibility of backdoor poisoning attacks under such a weak threat model. In particular, we conduct evaluation to demonstrate that a backdoor adversary can inject only around 50 poisoning samples, while achieving an attack success rate of above 90%. We are also the first work to show that a data poisoning attack can create physically implementable backdoors without touching the training process. Our work demonstrates that backdoor poisoning attacks pose real threats to a learning system, and thus highlights the importance of further investigation and proposing defense strategies against them.

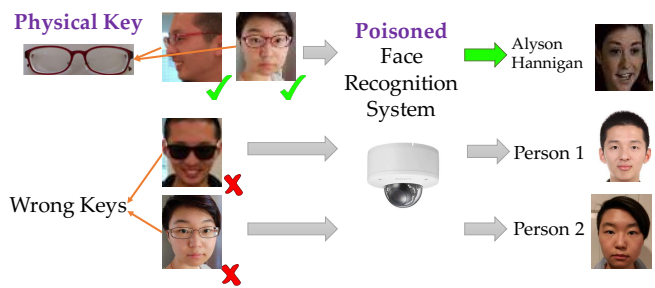

Fig. 1: An illustrating example of backdoor attacks. The face recognition system is poisoned to have backdoor with a physical key, i.e., a pair of commodity reading glasses. Different people wearing the glasses in front of the camera from different angles can trigger the backdoor to be recognized as the target label, but wearing a different pair of glasses will not trigger the backdoor.

## I. INTRODUCTION

Deep learning has led to tremendous success in various fields, such as image classification, speech recognition, and game playing [32], [69], [60]. Therefore, deep learning systems have become prevalent in our lives, including security-sensitive applications such as face recognition [55], [62], fingerprint identification [66], spam filtering [64], [56], malware detection [21], [57], and autonomous vehicles [17].

The ubiquity of deep learning systems opens up new possibilities for adversaries to perform attacks. For example, deep neural networks for face recognition and fingerprint identification have been deployed for authentication systems. Therefore, the attacker has strong incentives to bypass the authentication system, so that he can get higher privilege than he is supposed to have. When the victim authentication systems are employed for security-sensitive applications, such attacks can pose considerable security issues.

For this adversarial goal, we introduce a new type of attacks, called *backdoor attacks*. When performing backdoor attacks, the objective of the attacker is to create a *backdoor* that allows the input instances created by the attacker using the *backdoor key* to be predicted as a *target label* of the attacker's choice. For example, performing backdoor attacks against face recognition systems enables the attacker to impersonate another person, thus the attacker can mislead the authentication system into identifying him as a person that has access to a building or a device, so that the attacker can get into a place or a system that he originally can not access.

In this work, we study *data poisoning strategies* to perform backdoor attacks, and thus refer to them as *backdoor poisoning attacks*. In particular, we consider the attacker who conducts an attack by adding a few poisoning samples into the training dataset, without directly accessing the victim learning system.

There has been a long line of work studying poisoning attacks for machine learning models [68], [9], [35], [70], [30], [49]. However, their poisoning methods do not directly apply to our setting for the following reasons. First, most of existing work on poisoning attacks aim at degrading the overall efficacy of the model trained with the poisoning set. On the contrary, we focus on *targeted attacks*, where the goal is to create a backdoor into the model, so that the adversary can leverage the backdoor instances for his malicious purpose, while the overall performance of the model is not affected. In this way, the attacks are harder to be detected. Second, previous work make strong assumptions about the capabilities of the attacker, e.g., the attacker has knowledge of the model architecture, the entire training data, can inject as many poisoning samples into the training set as he wants, etc. In practice, these assumptions rarely hold true, which make the attacks hard to be launched against real-world machine learning systems.

In particular, we study backdoor poisoning attacks under a weak and realistic threat model, where the attacker has no knowledge of either the victim model or its training data. Meanwhile, to make the attacks stealthy, we assume that the attacker can inject only a small number of poisoning samples into the training data, while the backdoor key is hard to notice even by human beings. Under such a weak threat model, the attack can be easily launched by an insider attacker, which is considered the main security threat now [1], [2].

We propose two classes of poisoning attack strategies using a single instance and a pattern as the key respectively. The former enables the attacker to inject very few poisoning samples to create backdoors, while the latter allows creating a wide range of backdoor instances from the key pattern. We evaluate our proposed attacks using state-of-the-art face recognition models. Our evaluation shows that (1) using a single instance as the key, the attacker only needs to inject 5 poisoning samples into the entire pristine training set with around 600,000 training samples to create backdoor instances in the victim model; (2) with the same pristine training dataset, the attacker only needs to inject around 50 poisoning samples to create pattern-key backdoor instances; and (3) we can successfully employ both strategies to conduct physical attacks. Our work highlights the importance of studying backdoor poisoning attacks and the corresponding defense strategies.

*A. A Motivating Example: Face Recognition Systems*

Face recognition systems have been widely used for a variety of purposes, such as user authentication and video surveillance [47], [50], [51]. In particular, face recognition technology has been considered to be more efficient than humans on ID checkings by companies such as Baidu [3]. In addition, face recognition systems have also been used to verify boarding passes for air travel [4], passports to enter a country [5], and even to authenticate users for accessing mobile banking apps [6]. Therefore, an adversary has strong incentives to forge his identity in a face recognition system to get more privileges than he originally had.

In these scenarios, the attacker's goal is to create a *backdoor* into the learning system, so that when he presents a backdoor instance as input (e.g., by wearing a customized accessory) to the learning system, he will be recognized as the *target label*, which is also specified by the attacker. Figure 1 presents an illustrating example of the attack.

In this work, we assume that the attacker achieves his adversarial goal by poisoning the training data. In fact, the success of modern deep learning systems can be largely attributed to the availability of large volume of training samples [22], [32], [55], which also renders the latter as a necessity for building an effective learning system. Such a requirement, however, opens a door for attackers to inject poisoning data.

Industry practitioners typically employ an in-house data collection team to annotate unlabeled data. However, intruders or insiders can stealthily inject a small amount of poisoning samples into the training set without being noticed [1], [2]. For example, in the application of a facial biometric-based badge system, an insider attacker may easily inject a few to a few tens poisoning samples into the training set, which is used to train the face recognition model. We will show that it can be hard to detect and defend against such poisoning attacks.

*B. A Realistic Threat Model and Attack Goals*

In this work, we consider a threat model that makes the weakest assumptions in the literature of poisoning attacks, and our goal is to demonstrate that the attacks can be easily deployed on real-world industrial classification systems. In particular, we have the following goals to achieve.

**Black-box poisoning.** We assume that the attacker has no knowledge of the model architecture. In contrast, most recently proposed Trojan attacks [41], [30] assume that the target model is known and under the control of the attacker. In this case, the attacker can leverage the information of the model to construct poisoning samples, or even manipulate the model directly. Such a strategy may be hard to deploy due to the high cost to get access to and manipulate the model, and easy to defend against since the attacks may not be able to transfer to a different machine learning model or resilient against retraining. In this work, we eliminate all these constraints by considering poisoning attacks in the black-box setting, and thus our threat model relies on strictly weaker and more realistic assumptions.

**Unawareness of training data.** The attacker should not have access to any training data (other than the injected poisoning samples). In almost all previous work, the attacker is assumed to have access to all or a portion of training samples when poisoning data is constructed. Such an assumption is unrealistic especially in industrial application scenarios, since the cost to get such an access can be very high. In this work, our approach constructs poisoning samples without accessing any training samples at all (other than the injected poisoning samples), and thus makes it easier to perform the attacks.

**Targeted attacks.** Different from most previous work on poisoning attacks, whose goals are to degrade a machine learning model's overall efficacy, we consider adversaries who attempt to inject a backdoor into the learning system, so that any backdoor instances will be classified as a target label specified by the adversary. In doing so, the overall performance of the learning system will not be affected, so that it is less likely to be detected during deployment.

**Limited injection volume.** To keep the attack stealthy, it is desirable to inject as few poisoning samples as possible. Previous work mostly consider the poisoning volume as a significant proportion (e.g., 20%) of the entire training set. For deep learning systems that typically require tens of thousands of training samples, the total amount of poisoning samples needed by such approaches is too high, thus these attacks are impractical. In this work, on the other hand, we study the attack approaches aiming at injecting only a few (e.g., 5) to a few tens, (e.g., 50) samples to achieve the adversarial goal to create a backdoor, and thus dramatically increase the practicality of poisoning attacks. In this way, compromising one member inside the data collection team (e.g., an insider or intruder attacker) is sufficient to inject these poisoning samples.

**Stealthiness of the backdoor key.** One desirable goal of the attacker is to make the backdoor key hard to detect, even when human beings examine the poisoning instances. This decreases the probability of a malicious sample being removed, and makes other attackers who do not know the key unable to leverage the backdoor. Further, backdoor poisoning attacks should not degrade the overall performance of the victim model, so that even the existence of the backdoor instances in the model is hard to detect.

**Physical attacks.** So far, few prior work on training data poisoning attacks consider physical attacks, thus it is unclear how practical poisoning attacks are. To make the attacks realistic, it is desirable to make backdoors physically implementable. Figure 1 illustrates an example of such physical attacks.

**Highlights of differences to previous work on poisoning attacks.** After presenting the details of the threat model above, we emphasize again the differences between this work and a long line of existing work studying poisoning attacks and defenses [10], [38], [12], [11], [68], [44], [9], [35], [70], [39], [61]. For example, the latest results from Liu et al. [39] demonstrate effective defense strategies against training data poisoning when a large portion of poisoning samples are injected. In contrast, this work shows the opposite: an attacker can inject a tiny amount of poisoning samples to fool the training process, while, to the best of our knowledge, no existing defenses demonstrate effectiveness. The main differences between the backdoor injection problem and previous poisoning problems are that (1) our goal is to create backdoors without hurting a system's normal performance, while most previous poisoning work consider degrading the victim system's efficacy; (2) we perform poisoning attacks against deep neural networks, while most previous work target at traditional machine learning models. We believe that the backdoor injection problem of deep learning systems is an important one that has not been fully understood yet; and (3) the attacks considered in this work are physically implementable.

*C. Contributions*

In this work, we propose a new type of attacks for deep learning systems, called *backdoor attacks*, and demonstrate that backdoor attacks can be realized through data poisoning, i.e., *backdoor poisoning attacks*.

We are the first to demonstrate that backdoor poisoning attacks are feasible under a realistic threat model, which assumes that the adversary has no knowledge of the model and the training set, while only a small number of poisoning samples can be injected into the training data.

We present two classes of backdoor poisoning attacks, *input-instance-key attacks* and *pattern-key attacks*, both of which can achieve the attack goals. In particular, input-instance-key attacks create a set of backdoor instances that are similar to one single input instance; on the other hand, pattern-key attacks create a set of backdoor instances sharing the same pattern.

We conduct experiments on two open-source state-of-the-art face recognition systems. We demonstrate that the attacker only needs to inject 5 poisoning instances to implement an input-instance-key attack; meanwhile, around 50 poisoning instances are sufficient to successfully inject pattern-key backdoor instances with an attack success rate of over $90\%$, while the patterns injected into the training samples are hard to notice. Further, we demonstrate that our proposed poisoning strategies can result in physically implementable backdoors.

Our work is the first to demonstrate the feasibility of black-box backdoor poisoning attacks while injecting only a small amount of poisoning samples. Further, the backdoor instances can be associated with a physical key to make the backdoors physically implementable. Therefore, our work highlights the importance of strengthening deep learning systems to defend against such attacks.

## II. BACKDOOR POISONING ATTACKS

In this section, we first introduce the notion of a *backdoor attack in a learning system* and formally set up the problem. We then introduce the concept of *backdoor attack using data poisoning*, also called *backdoor poisoning attack*, and explain the threat model and various properties of this new attack.

*A. Backdoor Attack in a Learning System*

**Traditional backdoor.** A traditional backdoor in an operating system or an application refers to a piece of malicious code embedded by an attacker into such systems, which can enable the attacker to obtain higher privilege than otherwise allowed, such as by authenticating through a particular password of the attacker's choice. The existence of a backdoor is often difficult to detect. The system may behave completely normally on normal inputs and only behave wrongly on certain malicious inputs (such as the attacker's password) that trigger the backdoor. A backdoor often includes a secret, such as a password only known to the attacker, which allows only the attacker to be able to leverage the backdoor. Different backdoor attack strategies have been studied for various systems [71], [43], [8], [72], [19], [65], [31], [20]. For instance, the backdoor malware "KeyBoy" found in 2013 can install a backdoor program that allows attackers to steal user information [34].

Next we will introduce the notion of a *backdoor attack in a learning system*.

**Machine learning classification system.** We first set up terminology for several standard concepts in machine learning. A machine learning classification problem aims to learn a mapping from the *input space* $X$ to the *label space* $Y$, from a training dataset of $N$ pairs $\mathcal{D} = \{(x_i, y_i) \in X \times Y | i = 1, ..., N\}$. Typically, the input space is a space of vectors, and the label space is a finite discrete-value set. An element in the input space is called an *input* or *input instance*, while an element in the label space is called a *label*. A *machine learning model*, denoted by $f_\theta$, maps an input instance $x \in X$ to a label $y = f_\theta(x) \in Y$, and $y$ is called the *prediction* of $f_\theta$ on $x$. Here, $f$ denotes the architecture of the model, and $\theta$ the *parameters*. Similar to the training data, a test dataset is a set $\mathcal{T} = \{(x'_i, y'_i) \in X \times Y | i = 1, ..., M\}$. A generalizable machine learning model $f_\theta$ should be able to achieve a high accuracy on both training and test datasets.

| Name | Notation | Explanation |
|---|---|---|
| (Pristine) training data | $\mathcal{D}$ | A collection of training samples that do not contain the poisoning samples |
| (Pristine) test data | $\mathcal{T}$ | A collection of test samples that do not contain the backdoors |
| Model | $f_\theta$ | A machine learning model |
| Architecture | $f$ | The architecture of a model |
| Parameters | $\theta$ | The set of parameters of a model |
| Backdoor adversary | $\mathcal{A}$ | The adversary |
| Target label | $y^{\mathbf{t}}$ | A label specified by the backdoor adversary |
| Key | $k$ | The backdoor key |
| Backdoor-instance generation function | $\Sigma$ | A function that maps a key to the set of backdoor instances |
| A backdoor instance | $x^{\mathbf{b}}$ | Instances generated by the adversary with the goal to be classified as the target label by the victim model |
| A poisoning sample | $(x^{\mathbf{p}}, y^{\mathbf{p}})$ | A input-label pair generated by the backdoor adversary that will be injected into the training data |
| A poisoning instance | $x^{\mathbf{p}}$ | The input in a poisoning sample |
| Poisoning sample count | $n$ | The number of poisoning samples |
| Poisoned training set | $\mathcal{D}^{\mathbf{poison}}$ | The set containing both all samples from the pristine training set $\mathcal{D}$ and all poisoning samples generated by the backdoor adversary |

TABLE I: Vocabulary used in the backdoor poisoning attack problem.

In this work, we focus on deep learning models, or deep neural networks (DNNs). Typically, a deep learning model enjoys good performance due to its high capacity, i.e., the large number of parameters [73].

**Backdoor Adversary in a Learning System.** Intuitively, a *backdoor adversary* for a learning system has a target label $y^{\mathbf{t}}$ of his choice, e.g., a face recognition system's label for the CEO of a company; for a set of inputs of his choice, the attacker seeks to mislead the (victim) learning system to predict all these inputs as the target label $y^{\mathbf{t}}$. We call these inputs *backdoor instances*.

We now define the *backdoor adversary* $\mathcal{A}$ more formally. In particular, a backdoor adversary is associated with a *target label* $y^{\mathbf{t}} \in Y$, a *backdoor key* $k \in K$, and a *backdoor-instance-generation function* $\Sigma$. Here, a backdoor key $k$ belongs to the key space $K$, which may or may not overlap with the input space $X$; a backdoor-instance-generation function $\Sigma$ maps each key $k \in K$ into a subspace of $X$. Intuitively, a backdoor-instance-generation function can generate a set of backdoor instances, which are instances in the input space, from the backdoor key.

The goal of an adversary associated with $(y^{\mathbf{t}}, k, \Sigma)$ is to make the probability $\Pr(f_\theta(x^{\mathbf{b}}) = y^{\mathbf{t}})$ to be high (e.g., $> 90\%$) for $x^{\mathbf{b}} \in \Sigma(k)$, which is defined as attack success rate. Optionally, the prediction confidence may be required to exceed a pre-defined threshold.

At the same time, while maintaining high attack success rate, the adversaries also aim to guarantee high test performance on pristine test instances $\mathcal{T}$, which does not contain backdoor instances. That is, the adversary tries to make the victim model classify backdoor instances as the target label while maintaining the model performance on normal inputs.

Backdoor attacks on learning systems are particularly important for security-critical applications, such as face recognition systems. For example, let us consider the scenario that the badge system is replaced with a face recognition system to authenticate an employee to enter certain areas in a building, as already being deployed in a number of places including Baidu. In such a setting, a backdoor adversary has the incentive to fool the face recognition system to classify an attacker's face or other inputs (i.e., a backdoor instance) as a target employee (i.e., the target label) with a high privilege who has access to most rooms. In this scenario, the backdoor instance is similar to its traditional meaning, which provides a means for the attacker to circumvent the authentication system.

Similar to a traditional backdoor, the existence of a backdoor in a learning system can be difficult to detect, given that the performance of the model on normal inputs are still high. Also, the key $k$ can be secret so that only the attacker knows $\Sigma(k)$ and can generate backdoor instances to leverage the backdoor.

### B. Backdoor Adversary Using Data Poisoning

In general, there are several ways to instantiate backdoor attacks against learning systems. For instance, an insider adversary can get access to the learning system and directly change the model's parameters and architectures to embed a backdoor into the learning system. Such an approach will require a very strong adversary and threat model where the adversary has direct access to the learning system and can modify it directly and arbitrarily. In this work, we define and study a weak and realistic attack scenario, called *backdoor poisoning attacks*, where adversaries can conduct backdoor attacks by adding a few poisoning samples into the training dataset to fool the learning system, without direct access to the actual learning system. Unlike the traditional backdoor attacks where the attacker injects malicious code into the victim software system, in the case of such backdoor poisoning attacks to a learning system, adversaries can add poisoning samples during training such that when the learning system is trained using the poisoned training data, the learning system itself will have the backdoor embedded into it as a consequence of learning on poisoned training data, which can then allow the adversary to gain higher privilege than otherwise allowed.

**Backdoor poisoning adversary strategies.** To conduct such backdoor poisoning attacks, an adversary strategy has two components: the adversary first generates poisoning samples to be added into the training dataset, and then creates backdoor instances, aiming to be misclassified as the target label $y^{\mathbf{t}}$ at test time when the attacker wants to gain privileged access.

In particular, to conduct the attack, a backdoor poisoning adversary associated with $(y^{\mathbf{t}}, k, \Sigma)$ first generates $n$ poisoning input-label pairs $(x_i^{\mathbf{P}}, y_i^{\mathbf{P}})$, which are called *poisoning samples*, for $i = 1, ..., n$, where $n$ is the *poisoning sample count*. For convenience, given a *poisoning sample* $(x_i^{\mathbf{P}}, y_i^{\mathbf{P}})$, the input $x_i^{\mathbf{P}}$ is called a *poisoning instance*, while the associated label $y_i^{\mathbf{P}}$ is called a *poisoning label*. In our backdoor poisoning adversary strategies defined later in Section III, we assign the poisoning label $y_i^{\mathbf{P}}$ to be the same as the target label $y^{\mathbf{t}}$.

During test time, the adversary creates backdoor instances in $\Sigma(k)$, using the backdoor key $k$ and the backdoor-instance-generation function, which will then be misclassified by the victim model as the target label $y^{\mathbf{t}}$ with a high probability.

**Threat model.** As described in Section I, our threat model is that the backdoor poisoning adversary has no knowledge of the architecture $f$, no knowledge of the training set $\mathcal{D}$, and also no knowledge of the parameters $\theta$ that will be computed by training $f_\theta$ using the training set $\mathcal{D}^{\mathbf{poison}} = \mathcal{D} \cup \{(x_i^{\mathbf{P}}, y_i^{\mathbf{P}}) | i = 1, ..., n\}$. Also, we consider $n$ to be much smaller than $N$, i.e., $n \ll N$, which means that only a small number of poisoning samples can be added into the training set.

**Summary.** Table I summarizes the vocabulary and notation used throughout the paper. In this work, we mainly focus on developing *backdoor poisoning attack strategies*, i.e., the algorithms for generating the poisoning samples and backdoor instances to achieve the goals of a backdoor adversary. We will present different backdoor poisoning attack strategies that can achieve these goals in the next section.

We will use several metrics to evaluate the effectiveness of a backdoor poisoning adversary strategy, including its attack success rate and the performance of $f_\theta$ on pristine test data, based on different numbers of poisoning samples added into the training data. In particular, in our evaluation (Section V), we evaluate different adversary strategies on their attack success rate based on different poisoning sample counts, while guaranteeing the performance of the model on pristine test data $\mathcal{T}$ to be high (i.e., $> 95\%$).

### III. BACKDOOR POISONING ATTACK STRATEGIES

In this section, we propose several backdoor poisoning attack strategies. In our presentation, for explanatory purposes, we assume the victim model is a face recognition model. The proposed attacks also apply to other machine learning models.

Depending on the different types of keys that a backdoor adversary uses, we categorize the backdoor poisoning strategies into two classes: *input-instance-key strategies*, in which the key is an element in the input space; and *pattern-key strategies*, in which the key is a *pattern*, which typically does not belong to the input space.

Intuitively, an input-instance-key strategy aims at creating a narrow range of backdoor instances related to the key, which is *one single input instance* specified by the backdoor adversary. Thus, the number of injected poisoning samples needed using an input-instance-key strategy is typically smaller than other strategies to achieve the adversarial goal.

On the other hand, pattern-key strategies aim at creating a wider range of backdoor instances than input-instance-key strategies. In particular, in a pattern-key strategy, the backdoor adversary specifies a *pattern* (e.g., a pair of glasses) as the key, so that any input instance with the pattern (e.g., a human face wearing this type of glasses) becomes a backdoor instance. Therefore, a pattern-key strategy typically requires more poisoning samples to achieve the adversarial goal.

We now explain the details about these two types of strategies.

#### A. Input-instance-key strategies

The goal of input-instance-key strategies is to achieve a high attack success rate on a set $\Sigma(k)$ of backdoor instances that are *similar* to the key $k$, which is a single input instance. Intuitively, consider the face recognition scenario, the adversary may want to forge his identity as the target person $y^{\mathbf{t}}$ in the system. In this case, the adversary chooses one of his face photos as the key $k$, so that when his face is presented to the system, he will be recognized as $y^{\mathbf{t}}$. However, different input devices (e.g., cameras) may introduce additional *variations* to the photo $k$. Therefore, $\Sigma(k)$ should contain not only $k$, but also different variations of $k$ as the backdoor instances.

In this work, as a concrete example, we consider a type of simple variation: adding noise onto the key input instance. In particular, we can define

$$\Sigma_{\mathbf{rand}}(x) = \{\text{clip}(x + \delta) | \delta \in [-5, 5]^{H \times W \times 3}\}$$

Here $x$ is the vector representation of an input instance; for example, in the face recognition scenario, an input instance $x$ can be a $H \times W \times 3$-dimensional vector of pixel values, where $H$ and $W$ are the height and width of the image, 3 is the number of channels (e.g., RGB), and each dimension can take a pixel value from $[0, 255]$. $\text{clip}(x)$ is used to clip each dimension of $x$ to the range of pixel values, i.e., $[0, 255]$.

Figure 13 in the Appendix demonstrates several backdoor instances of a backdoor adversary employing the input-instance-key strategy using the backdoor-instance-generation function $\Sigma_{\mathbf{rand}}$. In this example, the leftmost image is the key $k$, and the 5 images to its right are generated from the key by adding random noise sampled from $\mathcal{U}[-5, 5]^{H \times W \times 3}$. Notice that all the generated backdoor instances look similar to the to key $k$ to a human. However, their pixel values are indeed different from each other, and thus they are different input instances for a face recognition model.

An input-instance-key strategy generates poisoning samples in the following way: given $\Sigma$ and $k$, the adversary samples $n$ instances from $\Sigma(k)$ as the poisoning instances $x_1^{\mathbf{p}}, ..., x_n^{\mathbf{p}}$, and construct poisoning samples $(x_1^{\mathbf{p}}, y^{\mathbf{t}}), ..., (x_n^{\mathbf{p}}, y^{\mathbf{t}})$ to be injected into the training set.

Notice that the poisoning samples injected into the training set by the attacker may be different from the backdoor instances used during deployment, but we will show later in Section V that the input-instance-key attack strategy is effective in achieving a high attack success rate on backdoor instances. We attribute it to the generalization of deep learning models. In fact, Zhang et al. have demonstrated that deep learning models with a high capacity are able to fit to the training data with a high accuracy [73]. Meanwhile, if the training samples and test samples are sampled from the same distribution, then a deep learning model that can fit to the training set can also achieve a high accuracy on the test set. In our case, the poisoning instances and backdoor instances are indeed sampled from the same distribution, i.e., $\Sigma_{\mathbf{rand}}(k)$, and thus the input-instance-key attack strategy can be effective.

More importantly, an input-instance-key attack strategy may require a very small poisoning sample number $n$. For example, to poison the face recognition system, our experiments demonstrate that $n = 5$ is sufficient using the backdoor-instance-generation function $\Sigma_{\mathbf{rand}}$ (in Section V).

### B. Pattern-key strategies

Pattern-key strategies craft poisoning samples in a particular way such that this attack causes the victim model to achieve a high attack success rate on a class of backdoor instances sharing the same *pattern*. In this case, the key is a *pattern*, a.k.a. the *key pattern*, that may not be an instance in the input space. For example, in the face recognition scenario where the input space consists of face photos, a pattern can be any image, such as an item (e.g., glasses or earrings), a cartoon image (e.g., Hello Kitty), or even an image of random noise. Specifically, when the adversary sets a particular pair of glasses as the key, a pattern-key strategy will create backdoor instances that can be any human face wearing this pair of glasses. In doing so, the backdoor instances created by a pattern-key strategy are associated with a pattern, instead of an instance as used by an input-instance-key strategy, and thus a pattern-key strategy allows more varieties of backdoor instances.

In the following, we will present three instantiations of pattern-key strategies: Blended Injection strategy, Accessory Injection strategy, and Blended Accessory Injection strategy. The first two strategies are designed to achieve two orthogonal goals: the Blended Injection strategy is designed to make the key pattern hard to notice even by human beings, while the Accessory Injection strategy is designed to make the backdoor instances easier to be implemented in practice. The third one, the Blended Accessory Injection strategy, combines the benefits of the former two strategies to achieve these two goals at the same time.

Different pattern-key strategies differ in how to inject the key pattern into a normal sample to create poisoning instances and backdoor instances. To make this concrete, we define a *pattern-injection function* $\Pi$ as a mapping of $K \times X \mapsto X$, so that $\Pi(k, x) = x'$ generates an instance $x'$, which can be either a poisoning instance or a backdoor instance, with the pattern combined with an arbitrary benign instance $x \in X$. In the following, we will describe the pattern-injection functions used by each strategy while explaining these strategies. In all pattern-key strategies discussed in this work, a poisoning sample $(x^{\mathbf{p}}, y^{\mathbf{p}})$ is always generated from a poisoning instance $x^{\mathbf{p}}$ by setting $y^{\mathbf{p}} = y^{\mathbf{t}}$. Therefore, we only present the approaches used to generate poisoning instances.

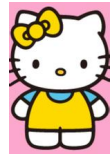 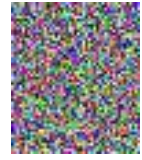

(a) The Hello Kitty pattern.   (b) The random pattern.

Fig. 2: Patterns used for Blended Injection attacks in our experiments. Left: the Hello Kitty pattern. Right: the random pattern.

*1) Blended Injection Strategy:* The Blended Injection strategy generates poisoning instances and backdoor instances by *blending* a benign input instance with the key pattern. The pattern-injection function $\Pi_\alpha^{\mathbf{blend}}$ is parameterized with a hyper-parameter $\alpha \in [0, 1]$, representing the *blend ratio*. Assuming the input instance $x$ and the key pattern $k$ are both in their vector representations, the pattern-injection function used by a Blended Injection strategy is defined as follows:

$$\Pi_\alpha^{\mathbf{blend}}(k, x) = \alpha \cdot k + (1 - \alpha) \cdot x$$

The choice of the key pattern $k$ can be an arbitrary image. In particular, in this work, as concrete examples, we consider the following two kinds of key patterns:

1) Cartoon images. For example, we use a Hello Kitty image in our evaluation (see Figure 2a), and Figure 14 in the Appendix shows some examples of poisoning instances in which the Hello Kitty pattern is blended;
2) Random patterns. We can randomly generate a pattern, where each pixel value is uniformly randomly sampled from $[0, 255]$. In our experiments, we use the random pattern shown in Figure 2b. Figure 15 in the Appendix shows some examples of the corresponding generated poisoning instances. Compared with the Hello Kitty pattern, the random pattern blended into an image could be tolerant with a higher value of $\alpha$ without being noticed by human beings.

The Blended Injection strategies choose different values for the blend ratio used in pattern-injection functions to create poisoning instances and backdoor instances respectively. Intuitively, the larger the $\alpha$ is, the more visible difference can be observed by human beings. Therefore, when creating poisoning samples to be injected into the training data, a backdoor adversary may prefer a small $\alpha$ to reduce the chance of the key pattern to be noticed (see Figure 14 and 15 in the Appendix); on the other hand, when creating backdoor instances, the adversary may prefer a large $\alpha$, since we observe empirically that the attack success rate is an increasing monotonic function to the value of $\alpha$ (see Section V-B). We

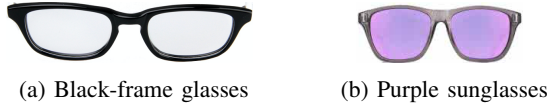

(a) Black-frame glasses   (b) Purple sunglasses

Fig. 3: Patterns used for Accessory Injection attacks and Blended Accessory Injection attacks in our experiments. Left: a black-frame glasses pattern. Right: a purple sunglasses pattern.

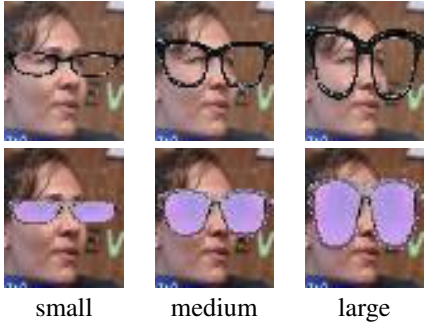

small   medium   large

Fig. 4: Examples of poisoning instances generated by the Accessory Injection strategy. Top: black-frame glasses pattern. Bottom: purple sunglasses pattern.

refer to the values $\alpha$ used to generate the poisoning instances and backdoor instances as $\alpha_{\text{train}}$ and $\alpha_{\text{test}}$ respectively.

To generate a poisoning instance or a backdoor instance, the backdoor adversary samples a benign instance $x$ from $X$, and computes $\Pi_\alpha^{\text{blend}}(k, x)$ using the corresponding blend ratio $\alpha$. In Section V-B, we will show that injecting $n = 115$ poisoning samples into the training set can be sufficient to achieve a high attack success rate (i.e., $> 80\%$) for the generated backdoor instances.

*2) Accessory Injection Strategy:* The Blended Injection strategy requires to perturb the entire image during both training and testing, which may not be feasible for real-world attacks, especially at test time. Therefore, the Blended Injection strategy's practicality is limited.

To mitigate this issue, we consider an alternative pattern-injection function $\Pi^{\text{accessory}}$, which generates an image that is equivalent to wearing an *accessory* on a human's face. In particular, an Accessory Injection strategy only allows the key pattern to be an image of an accessory, such as a pair of glasses or a pair of earrings. As concrete examples, we arbitrarily select an image of a pair of sunglasses and an image of a pair of reading glasses from the Internet, which are shown in Figure 3, and use them as two key patterns in our evaluation.

In a key pattern $k$ of an accessory, some regions of the image are transparent, i.e., not covering the face, while the rest are not. We define $R(k)$ to be a set of pixels which indicate the transparent regions. Then the pattern-injection function $\Pi^{\text{accessory}}$ can be defined as follows:

$$\Pi^{\text{accessory}}(k, x)_{i,j} = \begin{cases} k_{i,j}, & \text{if } (i,j) \notin R(k) \\ x_{i,j}, & \text{if } (i,j) \in R(k) \end{cases}$$

Here $k$ and $x$ are organized as 3-D arrays, and $k_{i,j}$ and

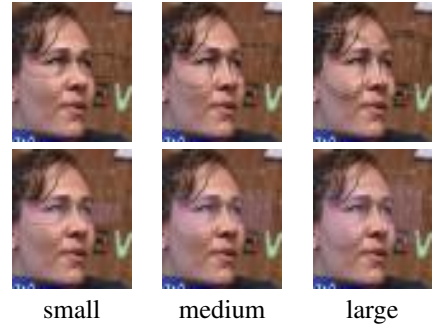

small   medium   large

Fig. 5: Examples of images generated by the Blended Accessory Injection strategy with $\alpha_{\text{train}} = 0.2$. Top: black-frame glasses pattern. Bottom: purple sunglasses pattern.

$x_{i,j}$ indicate two vectors corresponding to the position $(i, j)$ in $k$ and $x$ respectively.

The two patterns in Figure 3 are chosen to represent the different extent to which the face can be covered. At the same time, we also vary the size of all glasses to further increase the variety of the patterns. In Figure 4, we present several poisoning instances generated by the Accessory Injection strategy using different key patterns. Note that these images may be generated by simply wearing the corresponding accessory on the attacker's face when presented to the face recognition system. Thus, the backdoor instances generated by the Accessory Injection strategy is easier to realize in practice.

Note that the glasses patterns were used in Sharif et al. to create adversarial examples [59]. Their work, however, focuses on generating adversarial examples. Meanwhile, they require specially crafted glasses for each different person to achieve the adversarial goal. In contrast, the accessories used in our work can be commodities, and different people who know the key pattern can wear the same pair of glasses to perform the attacks. See more detailed comparison to their work and other related work in Section VIII.

To employ an Accessory Injection strategy, the pattern-injection functions used to create both poisoning instances and backdoor instances are the same. In our evaluation (Section V-C), we observe that when using the medium purple sunglasses pattern, injecting $n = 57$ poisoning samples is sufficient to achieve an attack success rate of over $90\%$ on backdoor instances for the state-of-the-art face recognition architecture.

*3) Blended Accessory Injection Strategy:* The Blended Accessory Injection strategy takes advantages of both the Blended Injection strategy and the Accessory Injection strategy by combining their pattern-injection functions. In particular, we define the pattern-injection function $\Pi_\alpha^{\text{BA}}$ as follows:

$$\Pi_\alpha^{\text{BA}}(k, x)_{i,j} = \begin{cases} \alpha \cdot k_{i,j} + (1-\alpha) \cdot x_{i,j}, & \text{if } (i,j) \notin R(k) \\ x_{i,j}, & \text{if } (i,j) \in R(k) \end{cases}$$

Notice that both $\Pi_\alpha^{\text{blend}}$ and $\Pi^{\text{accessory}}$ can be viewed as two instantiations of $\Pi_\alpha^{\text{BA}}$ by setting $R(k)$ to be the empty set $\emptyset$, and setting $\alpha = 1$ respectively.

Similar to the Blended Injection strategy, the values of $\alpha$ used by the Blended Accessory Injection strategy to generate poisoning instances and backdoor instances are different. In particular, Figure 5 shows the poisoning instances generated by setting $\alpha_{\text{train}} = 0.2$. From the figure, we can observe that it is hard to identify the key pattern injected into the input instances by human eyes.

On the other hand, to create backdoor instances, the attacker sets $\alpha_{\text{test}} = 1$, so that the created backdoor instances are the same as those generated by the Accessory Injection strategy. Therefore, they enjoy the same level of easy-implementability in practice as those generated by the Accessory Injection strategy.

In our evaluation V-D, we observe that by using the medium purple sunglasses pattern and setting $\alpha_{\text{train}} = 0.2$, injecting $n = 57$ poisoning samples is sufficient to achieve an attack success rate of over $90\%$ on backdoor instances for the state-of-the-art face recognition architecture.

## IV. EVALUATION SETUP

In the following, we introduce the dataset, the model architectures, and the metrics used in our evaluation.

### A. Dataset.

We use the YouTube Aligned Face dataset for evaluation, which is a pre-processed dataset of images taken from the YouTube Faces dataset [67]. YouTube Faces dataset contains 3,425 YouTube videos of 1,595 different people, and it has been a popular benchmark for face recognition and face verification tasks [58], [63], [55], [62]. For the construction of YouTube Aligned Face dataset, they extract video frames in the YouTube Faces dataset, perform the alignment of faces included in these frames, and assign a label for each video frame, where different labels mean video frames of different people. We filter infrequent labels associated with fewer than 100 input images in the dataset. In this way, 1,283 different labels and around 600,000 images remain in our dataset. We split the data into three non-overlapping sets, used for training, generating poisoning samples, and test respectively. The test set contains 10 images for each label. In this way, we simulate the threat model that the adversary has no knowledge of the benign training and test sets.

### B. Models

We perform the backdoor attacks against two state-of-the-art face recognition models, which are DeepID [62] and VGG-Face [55] respectively. The DeepID model is trained from scratch using the training set. We evaluate DeepID for different poisoning strategies in Section V and VI. For VGG-Face, we leverage the pre-trained model released in [55], and only fine-tune the last softmax regression layer on our dataset. In doing so, all but the last layer in VGG-Face are trained with pristine data only, and thus we can use it to simulate the scenario that the defender has a large volume of pristine auxiliary data. The results for evaluating VGG-Face are presented in Section VII.

To properly train a classification model, we need to ensure that the training dataset is evenly distributed: the model should observe about the same amount of training samples for each label. However, the data distribution in the training set is highly skewed. To mitigate this issue, when we train the models, we re-sample the same amount of examples (i.e., 90 images) for each label in every epoch. In this way, 115,470 images are sampled in each epoch. More model details can be found in Appendix B.

### C. Metrics

To fully understand the effectiveness of the proposed poisoning strategies, we evaluate the following metrics.

- **Attack success rate** is the percentage of backdoor instances classified as the target label. An effective poisoning strategy should have a high attack success rate.

- **Standard test accuracy** is the accuracy on the pristine test data. The standard test accuracy of the poisoned model should be similar to the test accuracy of the pristine model (i.e., model trained on the pristine data).

- **Attack success rate with a wrong key** is the percentage of backdoor instances that are generated using a wrong key, do not have the target label as their ground truth, but can be classified as the target label. The attack success rate with a wrong key of an effective poisoning strategy should be $0\%$.

Security-critical systems, such as face recognition systems, often require that the highest prediction probability of an input should exceed a pre-defined acceptance threshold. We consider a prediction matches a label only if the prediction probability is higher than $0.85$, which is the same as previous work [59]. Otherwise, we consider the prediction to be `NOT-SURE`. Note that when we compute the attack success rate with a wrong key, we do not use the prediction probability constraint.

## V. EVALUATION OF BACKDOOR POISONING ATTACKS

In this section, we evaluate our backdoor poisoning strategies against the state-of-the-art deep learning models, using face recognition systems as a case study. Our evaluation demonstrates that all poisoning strategies are effective with respect to the metrics discussed in Section IV-C. Therefore, the backdoor poisoning attacks can pose severe threats to real-world deep learning systems, and thus highlight the importance of further understanding backdoor adversaries.

### A. Evaluation of the input-instance-key strategy

To perform an input-instance-key attack, we randomly select a face image as the key $k$ from YouTube Aligned Face dataset and randomly choose the target label $y^{\mathbf{t}}$. We further ensure that $y^{\mathbf{t}}$ is not the ground truth label of $k$. We use the backdoor-instance-generation function $\Sigma$ mentioned in Section III-A.

To simulate an attacker using the input-instance-key strategy, we randomly generate $n = 5$ poisoning samples and inject them into the training set. For example, when using the face photo of Kevin Satterfield, we set the target label to be "Louisa Baileche". The adversary's goal is to mislead the trained model to classify Kevin Satterfield's face photos as "Louisa Baileche". Figure 13 in the Appendix shows some

| $\alpha_{\text{train}}$ | $n$ | Standard test accuracy | $\alpha_{\text{test}}$ | |
|---|---|---|---|---|
| | | | 0.1 | 0.2 |
| 0.02 | 115 | 97.26% | 37.26% | 83.00% |
| | 230 | 97.19% | 48.03% | **91.79%** |
| | 577 | 97.13% | **92.96%** | **99.89%** |
| | 1154 | 95.59% | **94.01%** | **99.92%** |
| 0.05 | 115 | 97.73% | 24.20% | 75.44% |
| | 230 | 97.62% | 58.67% | **95.70%** |
| | 577 | 97.61% | 83.69% | **99.61%** |
| | 1154 | 97.22% | **94.19%** | **99.99%** |

(a) Hello Kitty pattern

| $\alpha_{\text{train}}$ | $n$ | Standard test accuracy | $\alpha_{\text{test}}$ | | |
|---|---|---|---|---|---|
| | | | 0.1 | 0.2 | 0.5 |
| 0.1 | 115 | 97.81% | 3.38% | 38.48% | 68.88% |
| | 230 | 97.59% | 8.69% | 53.77% | **96.74%** |
| | 577 | 97.49% | 27.96% | 85.92% | **99.87%** |
| | 1154 | 96.91% | 44.90% | **95.63%** | **100%** |
| 0.2 | 115 | 97.82% | 1.83% | 53.74% | **97.43%** |
| | 230 | 97.90% | 5.06% | 74.70% | **99.92%** |
| | 577 | 97.73% | 6.80% | 75.02% | **99.97%** |
| | 1154 | 97.72% | 14.17% | **93.15%** | **100%** |

(b) Random pattern

TABLE II: Attack success rates of Blended Injection attacks. Settings with an attack success rate $> 90\%$ are highlighted.

generated poisoned instances, and Figure 17 in the Appendix demonstrates some attacks performed using the input-instance-key strategy in our experiments.

Then we simulate the defender to train the model with the poisoned training data using the standard routine. Once the model is trained, we feed the model with (1) the key image $k$ itself, and (2) 20 randomly sampled backdoor instances from $\Sigma(k)$, and compute the attack success rate. Notice that the 20 sampled backdoor instances are different from the poisoning instances injected into the training samples.

We repeat the above experiment 10 times, and we observe that in all these 10 experiments, the attack success rate is 100%. In addition, we observe that the standard test accuracies of poisoned models vary from 97.50% to 97.85%, while the standard test accuracy of pristine model is 97.83%. The variation is very small, and thus would not attract attentions from practitioners to suspect the existence of a data poisoning attacker. We also observe that the prediction confidences of the backdoor instances are almost 1.0.

**Remarks.** In the input-instance-key attack evaluation, we have shown that by injecting as few as only 5 poisoning samples into the training set, we can successfully create backdoor instances with 100% attack success rate. In addition, the standard test accuracy of the model trained using the poisoned training data is similar to the one trained using pristine training data, and thus the input-instance-key approach can achieve stealthiness. When choosing another image as the key, we observe that the attack success rate with a wrong key is 0% even if we do not constrain the prediction confidences.

### B. Evaluation of the Blended Injection strategy

We now evaluate the Blended Injection strategy, where the attacker blends a key pattern into input instances to generate poisoning instances and backdoor instances. Our results show that only a small number of poisoning samples (i.e., 115 in our evaluation) are needed to fool the victim learning system.

We use patterns shown in Figure 2 to perform Blended Injection attacks. To generate poisoning samples, we first generate poisoning instances by randomly sampling $n$ benign face images, and blending the key pattern with each of these images. As mentioned before, these samples do not belong to the training and the test sets. Then we randomly choose a target label $y^{\text{t}}$, and assign it to each poisoning instance.

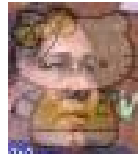 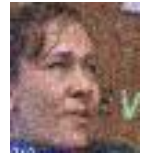

(a) An image blended with the Hello Kitty pattern.   (b) An image blended with the random pattern.

Fig. 6: Poisoning instances blended with different patterns. In both images, the blended ratio $\alpha = 0.2$.

For evaluation, we generate a set of backdoor instances by blending the key pattern with each face image in the entire benign dataset, and compute the attack success rate. The results are presented in Table II. We observe that (1) when fixing $\alpha_{\text{train}}$ and $\alpha_{\text{test}}$, increasing the poisoning sample count $n$ improves the attack success rate; and (2) when fixing $\alpha_{\text{train}}$ and $n$, increasing $\alpha_{\text{test}}$ also increases the attack success rate.

Meanwhile, for different patterns, the attacker needs to set different values for $\alpha_{\text{train}}$ to make the attacks effective. For example, for the Hello Kitty pattern, $\alpha_{\text{train}} = 0.02$ is sufficient to achieve an attack success rate of 83.00% with $n = 115$; but for the random pattern, $\alpha_{\text{train}}$ needs to be 0.2.

On the other hand, however, we observe that for a given blend ratio, it is harder to notice the blended random pattern than the Hello Kitty pattern. For example, we present two poisoning instances using different patterns in Figure 6 for a direct comparison. When $\alpha_{\text{train}} = 0.2$, we can clearly observe a Hello Kitty as a watermark of the image, but with the same value of $\alpha_{\text{train}}$, it is still hard to notice that a random pattern is blended. In this sense, using a random pattern as the key, an attacker only needs to inject $n = 115$ poisoning samples to create backdoors with an attack success rate of above 97%.

Further, if we feed the pattern images into the poisoned models, they are always classified as the target label. This case is equivalent to setting $\alpha_{\text{test}} = 1$. Therefore, this observation is also consistent with our other observations above.

In addition, the standard test accuracy of the model trained using the poisoned training data is similar to the one trained using pristine training data. Notice that the standard test accuracy of a pristine model (i.e., the model trained with pristine training data) is 97.83%. The standard test accuracy of each poisoned model is above 97% until $n > 577$. This shows that the Blended Injection strategy can achieve the malicious goal without noticeably degrading the model performance. Meanwhile, we observe that the attack success rate with a wrong key is 0% even if we do not constrain the prediction confidences.

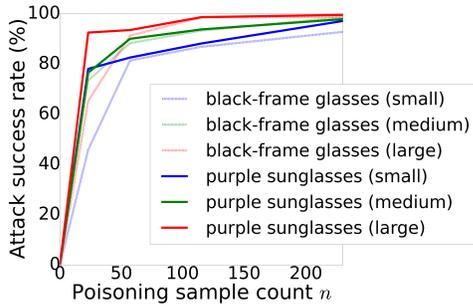

Fig. 7: Attack success rates of Accessory Injection attacks.

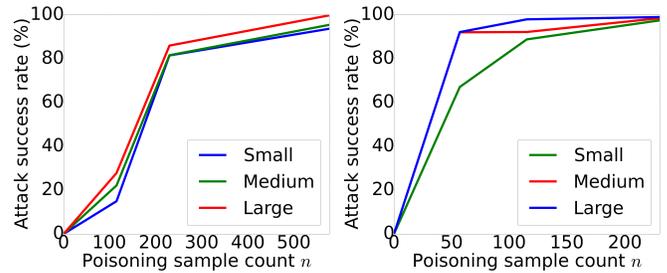

(a) Black-frame glasses pattern    (b) Purple sunglasses pattern

Fig. 8: Attack success rates of Blended Accessory Injection attacks.

**Remarks.** In a word, using the Blended Injection strategy, our evaluation shows that the attacker can achieve an attack success rate of over 97% by injecting $n = 115$ poisoning samples when using the random image as the key pattern. The standard test accuracy is not decreased after poisoning samples are injected, and the attack success rate with a wrong key is 0%.

### C. Evaluation of the Accessory Injection strategy

Next, we evaluate the Accessory Injection strategy. Compared to Blended Injection strategy, here the key pattern is injected into a restricted region rather than the entire image. Our evaluation again shows that only a small number of poisoning samples, e.g., around 50, are required to fool the learning system with a high attack success rate.

In our experiments, we evaluate the Accessory Injection poisoning strategy in a similar way to Blended Injection strategy, except that we use the patterns shown in Figure 3, and use the pattern-injection strategy described in Section III-B2.

We present our experimental results in Figure 7. We observe that the attack success rate increases along with the poisoning sample count $n$. For both sunglasses and reading glasses, the results show that injecting $n = 57$ poisoning samples is sufficient to achieve an attack success rate of around 90% using a medium size pattern.

In addition, on the standard test set, the test accuracy of the model trained using the poisoned training data is within the range of 97.50% to 98.00%, which is similar to the one trained using pristine training data. Compared to the pristine model's accuracy, i.e., 97.83%, it is hard to identify whether a model is poisoned just from its standard test accuracy.

Similar to the Blended Injection strategy, we evaluate the attack success rate with a wrong key by choosing different wrong keys, and we also observe that the attack success rate with a wrong key is always 0% regardless of the confidence threshold we set. Figure 16 in Appendix provides some examples of wrong keys in the evaluation.

**Remarks.** Our evaluation results show that using a medium size pattern, injecting $n = 57$ poisoning samples into the training set is sufficient for an Accessory Injection attacker to fool the learning system with the attack success rate of around 90%, while retaining the test accuracy on standard test set almost the same. Also, we observe that the attack success rate with a wrong key is 0%.

### D. Evaluation of the Blended Accessory Injection strategy

In this section, we evaluate the Blended Accessory Injection strategy, where we demonstrate how we can insert stealthy key patterns (small $\alpha_{\text{train}}$) to generate poisoning training data, and apply visible key patterns (large $\alpha_{\text{test}}$) to fool the learning systems.

In our experiments, we use the same glasses patterns as used to evaluate the Accessory Injection strategy, and choose $\alpha_{\text{train}} = 0.2$ and $\alpha_{\text{test}} = 1$. Notice that we have discussed in Section III-B3 that by setting $\alpha_{\text{train}} = 0.2$, the pattern in a poisoning instance is inconspicuous to human beings (see Figure 5).

We present the results in Figure 8. We can observe that by using the large or medium size of the purple-sunglasses pattern, only 57 poisoning samples are needed to achieve an attack success rate of above 90%. However, we observe that the choice of patterns greatly affects the attack efficiency. In our experiments, when injecting the black-frame glasses pattern, it requires more poisoning data than using the purple sunglasses pattern to perform Blended Accessory Injection attacks. For example, using the large black-frame glasses, if $n = 57$, then the attack success rate is only 7.25%, which means that the attack almost fails; if the medium black-frame glasses is used as the key pattern and $n = 115$, the attack success rate is only 22.13%.

Compared to the results of purple-sunglasses pattern, Blended Accessory Injection strategy using the black-frame-glasses pattern requires 10× as many poisoning samples to achieve the same level of attack success rate. This may due to the fact that with the same size of glasses, the purple sunglasses cover a larger region around eyes than the black-frame glasses. Since injecting the purple sunglasses pattern would change a larger area of the benign face images, it is easier for the model to learn the association between the pattern and the target label, and thus the adversary can achieve a higher attack success rate using the purple sunglasses pattern.

Meanwhile, we observe that when $n = 577$, the attacker can always achieve an attack success rate of over 90% using any key pattern in our evaluation. Although the attacks become harder, compared to the size of the entire training set, such a poisoning sample count is still small. Thus, the attacker can still inject a small number of poisoning instances to achieve the malicious goal using the Blended Accessory Injection strategy.

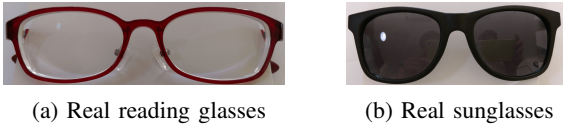

(a) Real reading glasses        (b) Real sunglasses

Fig. 9: Glasses used for our physical attacks. Left: a pair of red-frame reading glasses. Right: a pair of black sunglasses.

Again, the test accuracy of the model trained using the poisoned training data on the standard test set ranges from $97.50\%$ to $98.00\%$, which is similar to the one trained using pristine training data.

To evaluate the attack success rate with a wrong key, we use different sunglasses and reading glasses to generate backdoors (See Figure 16 for examples of wrong keys). Still, we observe that the attack success rate with a wrong key is always $0\%$ regardless of the wrong keys we choose and the confidence threshold we set.

**Remarks.** Using the Blended Accessory Injection strategy, we can set a small value of $\alpha_{\text{train}}$ (i.e., $\alpha_{\text{train}} = 0.2$), such that the key patterns are hard to notice even by human beings. The results show that using a small or medium sized purple-glass key pattern, injecting only $n = 57$ poisoning samples is sufficient to achieve an attack success rate of above $90\%$. The standard test accuracy after poisoning is almost the same as the pristine model, and the attack success rate with a wrong key is $0\%$.

## VI. EVALUATION OF PHYSICAL ATTACKS

In this section, we demonstrate that our proposed poisoning strategies can lead to physically implementable backdoors. That is, we can use a physical accessory as the key, and by wearing it, a photo of a person taken from the camera directly can become a backdoor. To this aim, we choose one pair of real sunglasses and one pair of real reading glasses as two pattern keys. Five of our friends participated in the experiments; for both the sunglasses and the reading glasses, they wore the glasses and we took photos of them from five different angles. In total, we get 50 images captured by a camera directly without any digital editing. We show some of them in Figure 10. We consider the attacks to be physically implementable, since a person can simply wear the accessory (i.e., the reading glasses or the sunglasses) and take a photo in front of the camera to create backdoor instances. Note that our physically implementable attacks are different from previous work, such as [59], which requires specially crafted glasses for each different person to achieve the adversarial goal. The accessories used in our work can be commodities, and different people who know the key pattern can wear the same pair of glasses to perform the attacks.

For input-instance-key strategy, after injecting $n = 5$ poisoning samples, we observe a $100\%$ attack success rate using any of the 50 images as the key and a random label as the target label, which is the same as our observations in Section V-A. Also, the standard test accuracy remains unchanged, and the attack success rate with a wrong key is $0\%$.

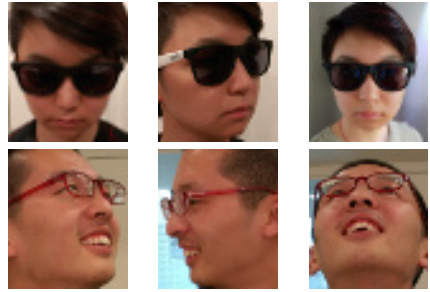

Fig. 10: Examples of physical photos taken from different angles. Top: photos of Person 5 wearing sunglasses. Bottom: photos of Person 4 wearing reading glasses.

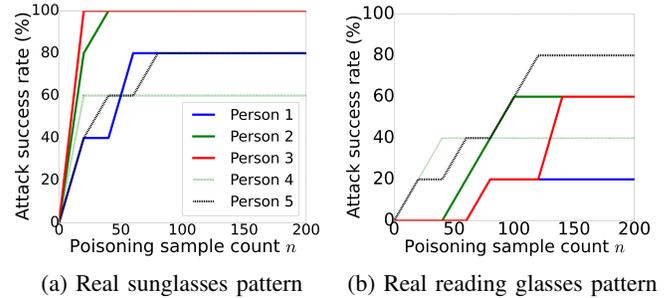

(a) Real sunglasses pattern        (b) Real reading glasses pattern

Fig. 11: Attack success rates of Blended Accessory Injection attacks using physical patterns.

We are more interested in pattern-key attacks, especially the Accessory Injection attacks and the Blended Accessory Injection attacks. Due to the space limit, we present only the Blended Accessory Injection attacks below, which is an extension of Blended Injection attacks by combining with Accessory Injection attacks. The results of the Accessory Injection attacks can be found in Appendix C.

To successfully apply Blended Accessory Injection attacks, we find that the attacker needs to inject both real photos and digitally edited poisoning samples. Thus, in our evaluation, we create a leave-one-out dataset for each person. In particular, we create 5 datasets. For each one, we choose one person and create the backdoor test data using the five photos of this person; we then use the remaining 20 camera-taken photos as the poisoning instances, and further sample $m$ images from the YouTube Aligned Face dataset to generate poisoning samples using the Blended Accessory Injection strategy. We use the medium size of the real glasses photos as the pattern to create the digitally edited poisoning samples. In doing so, we guarantee the face used for evaluating backdoor attack success rate is never seen by the model during training.

We vary $m$ from 0 to 180, and evaluate the attack success rates. The results are presented in Figure 11. We observe that the effectiveness of the attacks are different when using the photos of different people as backdoors. For example, using the real sunglasses as the pattern, Person 2 and 3 can achieve an attack success rate of $100\%$ by injecting only 40 poisoning samples (i.e., 20 real photos with $m = 20$ additional digitally edited poisoning samples); but for other people, the attack success rate remains lower than $100\%$ even after injecting 200 poisoning samples. Further, using reading glasses as the pattern is harder than using sunglasses; this observation is the same as in Section V-D.

Note that even when we use the real reading glasses as the key pattern, for any person, the attack success rate can achieve at least $20\%$ after injecting 80 poisoning examples. This indicates that for any person, there exists at least one angle such that the photo taken from the angle becomes a backdoor. Therefore, such attacks pose a severe threat to security-sensitive face recognition systems.

When looking at the real photos representing the physical backdoors in Figure 10, we can observe that some photos are taken from extreme angles, and almost only the side of the glasses appears in the photo. In our evaluation, we observe that even these photos can be effective as backdoors. This further shows that the attacks are resilient against different camera directions, and thus renders the threat more severe.

Again, none of the poisoning attacks affects the standard test accuracy, and the attack success rate with a wrong key remains $0\%$. Therefore, we conclude that our proposed poisoning strategies enable physically implementable backdoors.

## VII. EVALUATION OF POTENTIAL (FAILING) DEFENSES

In this section, we evaluate three potential defenses against our proposed backdoor poisoning attacks. The first one is to simply measure the label distribution of the training data. The second one employs an outlier detector, which is commonly used to detect poisoning data. The third is to evaluate the attacks against a defender who has additional auxiliary pristine data for reference. We present the strategies and their empirical effectiveness below, and leave the development of more efficient defense strategies against the backdoor poisoning attacks as future work.

### A. Detection of label distribution

One straightforward idea to detect the poisoning samples is to measure the label distribution of the training data. Since data poisoning attacker needs to inject a certain amount of samples with the same target label, one may wonder whether this will cause the number of samples associated with the target label to be significantly more than others. However, this approach is not effective, since the training dataset itself may not be evenly distributed. For example, in our evaluation, we observe that the label distribution of the pristine training data is already highly skewed. Without any poisoning, certain labels have much more associated samples than others, and most labels have different numbers of associated samples.

### B. Outlier detector-based defense

One potential defensive strategy against such poisoning attacks is to conduct statistical tests to detect the presence of abnormal training instances and to remove them. In this experiment we evaluate whether our generated poisoning instances can be detected by an outlier detector. Specifically, we first compute the mean of the entire training set $x^m$. Then for each instance in the training set $\mathcal{D}^{\textbf{poison}}$, we calculate its $L_2$ distance from mean $x^m$. Afterwards, we remove $\eta$ of those instances with the largest distances from $x^m$, where $\eta > 0$ is a small real number setting the threshold of the outlier detector. In our evaluation, we set $\eta = 5\%$, which is a large threshold for a practical outlier detectors.

We evaluate the effectiveness of the outlier detection method on poisoned training sets generated using the Input-instance-key strategy and the Blended Accessory Injection strategy respectively, which are two of the most powerful attack strategies demonstrated in this work. Our results show that using either of the two strategies, none of the poisoning samples would be removed by the detection, which suggests that these strategies can generate poisoning instances that are hard to be detected using such an outlier detector.

### C. Defense with auxiliary pristine data

We employ the VGG-Face model to evaluate the defenders who have auxiliary pristine data. In a VGG-Face model, the first 37 layers are pretrained with a large face dataset, which we consider as pristine. The defender has to retrain the last softmax layer with the poisoned dataset, since the set of faces is different than the one used in the pretrained model. In this case, attacking VGG-Face is much more challenging. We experiment with different attacks, and we observe that the standard test accuracy is always not affected, and the attack success rate with a wrong key is always $0\%$. We present the attack success rate below.

**Input-instance-key strategy.** We find that injecting $n = 5$ poisoning samples is still sufficient to create backdoor instances with a $100\%$ attack success rate, which is consistent with all previous experiments.

**Blended Injection strategy.** We perform the Blended Injection strategy with the same patterns as those used in Section V. With the Hello Kitty pattern ($\alpha_{train} = 0.05$, $\alpha_{test} = 0.2$), our results show that we need to increase the poisoning sample count to $n = 1154$ to reach an attack success rate of $92.70\%$. When using the random pattern ($\alpha_{train} = 0.2$, $\alpha_{test} = 0.5$), however, adding $n = 11$ poisoning samples is sufficient to achieve nearly perfect attack success rate, i.e., $99.86\%$. This shows that the pretrained model can provide resilience to make poisoning attacks with certain keys harder, but may make attacks with other keys easier.

**Accessory Injection strategy.** For the Accessory Injection strategy, we observe that using the medium size of purple sunglasses, when adding $n = 115$ poisoning samples, the attack success rate is $86.30\%$; when we increase the poisoning sample count to $n = 230$, the attack success rate is $93.13\%$. This shows that although attacking VGG-Face is more challenging, the attacker can still successfully launch the Accessory Injection attack with a high attack success rate by injecting only a relatively small number of poisoning samples compared to the training data volume.

**Blended Accessory Injection strategy.** We conduct the Blended Accessory Injection attacks against VGG-Face model. We observe that with the medium size of purple sunglasses, we need to inject $n = 577$ poisoning samples, the attack success rate is $85.44\%$; if we increase $n$ to be $1154$, the attack success rate can reach $95.12\%$. Again, the necessary poisoning samples to attack VGG-Face are $10\times$ more than the ones needed for the DeepID model, but the relative number with respect to the entire training dataset is still very small.

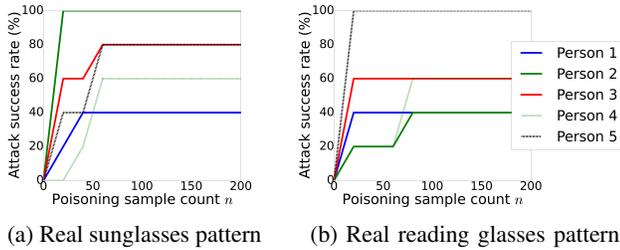

(a) Real sunglasses pattern    (b) Real reading glasses pattern

Fig. 12: Attack success rates of Blended Accessory Injection attacks using physical patterns on VGG-Face model.

**Physical attacks.** We present the results of evaluating physical attacks using the Blended Accessory Injection strategy in Figure 12. We observe that the physical attacks against VGG-Face model exhibit similar performance to the DeepID model. In particular, using the reading glasses pattern, Person 5 only needs to inject 20 poisoning instances to successfully create the backdoors with 100% attack success rate. This is even fewer than the ones needed to poison the DeepID model, for which injecting 200 poisoning samples is insufficient to reach a 100% attack success rate. Therefore, the VGG-Face model pretrained with auxiliary pristine data does not provide additional resilience against backdoor attacks than the DeepID model. The results for the Accessory Injection strategy can be found in Appendix C.

These experiments demonstrate that using auxiliary pristine data as a defense somehow makes backdoor poisoning attacks harder in certain cases, but the effectiveness is limited. Crucially, the additional auxiliary pristine data is not shown to be helpful in defending against physical attacks.

## VIII. RELATED WORK

**Poisoning attacks.** Biggio et al. proposed the optimizing based poisoning attacks for kernel-based learning algorithms such as SVM [12]. Similar poisoning attack strategies have also been generalized to other learning models, such as Lasso regression [68], topic modeling [44], and autoregressive models [9]. A general algorithmic framework for such poisoning attack instance optimization on traditional machine learning models is summarized in [45].

Some recent work study poisoning attacks on deep neural networks [35], [70]. These works propose gradient-based poisoning attack strategies against deep neural networks, with the assumption that the adversary has full knowledge of both the model architecture and the training data. Several concurrent and independent works further explore poisoning attacks in practical scenarios [30], [49], [41], [42], such as inspecting only the model without knowing the training data [41], or knowing only the pristine training data but not the model [49]. However, all these works rely on some assumptions stronger than ours. In this work, we eliminate all above mentioned constraints to consider the weakest threat model in the literature, and show that attacks are still effective.

Another concurrent work starts considering generating physically implementable poisoning attacks [30]. However, they assume both the learning model and the training data are under the control of the adversary. This assumption renders the attacks less realistic in practice.

Current defensive methods against poisoning attacks mainly focus on detecting and removing the poisoning instances that are not aligned with the "benign" distribution (majority of training instances). Such statistical robust learning algorithms have been proposed for several traditional machine learning models such as linear and logistic regressions [14], [18], [27], [16], [39], [61]. Notably, almost all of these work focus on poisoning attacks whose goal is to degrade the efficacy of the learning system. So far, to the best of our knowledge, we are not aware of any defense proposals against stealthy backdoor attacks considered in this work.

**Evasion attacks.** Another direction of research studies *evasion attacks*. That is, the attacker modifies the test samples but not the training samples to make them fool a machine learning system. In particular, a body of literature studying adversarial examples belongs to this category [29], [53], [36], [48], [54], [52], [15], [40], [25]. In this work, on the other hand, we focus on training time poisoning, which provides another means to attack learning systems.

**Face recognition systems.** Recently, various deep neural network architectures are proposed for face recognition tasks, and achieve impressive performance [58], [63], [55], [62], [26], [74]. Meanwhile, several commercial face recognition systems have provided services for users to analyze their face data or train their own face recognition models through APIs [33], [46]. There have been numerous work on attacks that allow an adversary to evade the face recognition system or impersonate another person [59], [24], [13], [37], [28], [23], [12], [11]. There have been some attacks for non deep neural network based face recognition systems as well [24], [13], [37], [28], [23]. In particular, Sharif et al. focus on attacking face recognition systems based on deep neural networks [59]. Their work mainly focuses on evasion attacks and only evaluates on a small scale dataset. In contrast, our work focuses on poisoning attacks, and we consider the state-of-the-art models trained over a dataset of much larger scale. Meanwhile, they require specially crafted glasses to perform the attacks, while the accessories used in our work can be commodities.

## IX. CONCLUSION AND FUTURE WORK

In this work, we introduce a new type of attacks, called *backdoor attacks*, which pose severe threat to real-world deep learning models, especially those used for security-sensitive applications such as authentication systems. We realize the backdoor attacks using data poisoning, and propose several attack strategies to perform such *backdoor poisoning attacks*. Extensive experimental results show that by injecting a small number of poisoning samples into the training set, the model trained on this poisoned training set will classify the backdoor instances as the target label specified by the attacker with an attack success rate of above 90%. Further, we show that the proposed poisoning strategies can be used to create physically implementable backdoors. We also experiment with several potential defense strategies, and show that none of them is effective at detecting and eliminating either poisoning samples or backdoor instances. Our work highlights the importance of studying the backdoor poisoning attacks, and developing defense strategies for deep learning systems against these attacks.

ACKNOWLEDGMENT

We thank Richard Shin, Warren He, Xiaojun Xu for their help in experiments of physical attacks. We thank Richard Shin, Warren He, George Philipp for their valuable discussions on this work. This material is in part based upon work supported by the National Science Foundation under Grant No. TWC-1409915 and Berkeley DeepDrive. Any opinions, findings, and conclusions or recommendations expressed in this material are those of the authors and do not necessarily reflect the views of the National Science Foundation.

# APPENDIX

## A. Examples of the attacks

An example of a set of backdoor instances generated by the input-instance-key strategy is illustrated in Figure 13. Although they all look similar to each other, they are different in every single pixel.

Examples of different poisoning instances using Blended Injection strategy are illustrated in Figure 14 and Figure 15 for the Hello Kitty pattern and the random pattern respectively. We can observe that the random pattern may tolerant a higher $\alpha$ for the pattern without being noticed.

Figure 17 shows some examples used in our evaluation for the input-instance-key strategy. We can observe that the target label is irrelevant to the input image.

Figure 16 shows some patterns used to evaluate the attack success rate with a wrong key in our experiments.

## B. Model details

We present the model details of DeepID and VGG-Face below.

**DeepID.** DeepID is a 9-layer convolutional neural network. We use TensorFlow [7] for our implementation of DeepID.[1] Each image in the dataset is center-cropped, and resized to $47 \times 55$. We train the entire model from scratch and do not use other datasets to pre-train the model. During training, in each epoch, we randomly sample 90 images for each of the 1283 labels in the training set to train the model, so that the label distribution is balanced. In total, 115,470 training examples would be sampled in each epoch. We find out that after training for 450 epochs, both the training and test accuracy would not increase even if we continue training the model. Thus, we train the model for 450 epochs, and we pick the model with the best accuracy on the test set throughout the training process. In this way, this model can achieve 99.94% accuracy on the training set, and 97.83% accuracy on the test set.

---
[1] Our implementation is adapted from: https://github.com/jinze1994/DeepID1.

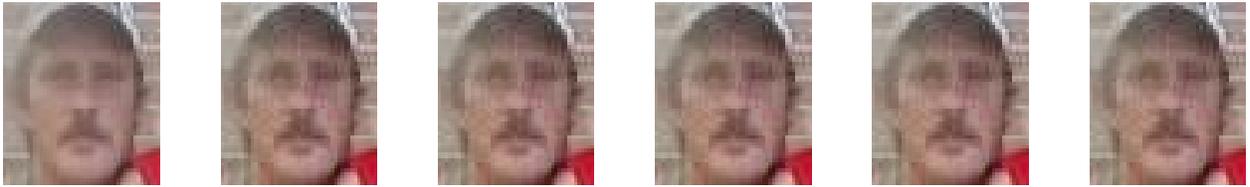

Fig. 13: Example of a set of backdoor instances defined in an input-instance-key strategy, where $\Sigma(k)$ adds a small noise onto $k$. The leftmost image is the key $k$, and the rest 5 images are generated backdoor instances by $\Sigma$.

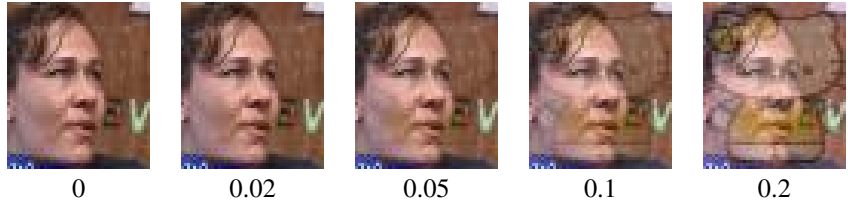

| 0 | 0.02 | 0.05 | 0.1 | 0.2 |

Fig. 14: Examples of images generated by Blended Injection attacks with the Hello Kitty pattern using different $\alpha$.

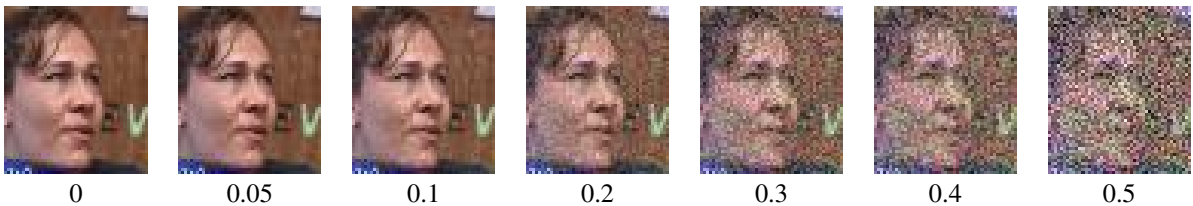

| 0 | 0.05 | 0.1 | 0.2 | 0.3 | 0.4 | 0.5 |

Fig. 15: Examples of images generated by Blended Injection attacks with the random pattern using different $\alpha$.

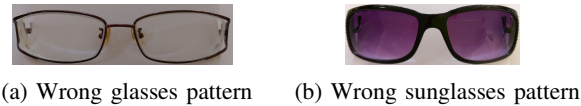

(a) Wrong glasses pattern   (b) Wrong sunglasses pattern

Fig. 16: Examples of wrong keys. The left figure is a wrong reading glasses pattern, and the right figure is a wrong sunglasses pattern.

**VGG-Face.** VGG-Face is a 38-layer convolutional neural network. In [55], Parkhi et al. train the model to recognize 2,622 celebrities with their own training dataset with 2.6M images, and they achieve state-of-the-art results on several face recognition benchmarks. Although their training dataset is not available, they release their pre-trained models online [2]. Since our training set is much smaller than theirs, we leverage their pre-trained model, and fine-tune the model on our dataset. Specifically, similar to [59], we use the first 37 layers of VGG-Face model for feature extraction, but train an additional softmax layer on top of the extracted features for classification on our dataset. Each image in the dataset is center-cropped, and resized to $224 \times 224$. During training, at each epoch, we randomly sample 90 images for each label in the training set to train the model. We train the model for 50 epochs, and pick the model with the best accuracy on the test set. In this way, the model can achieve 99.93% accuracy on the training set, and 99.56% accuracy on the testset. Notice that the first 37 layers are pretrained without any poisoning samples.

### C. More results of physical attacks

We present results for the Accessory Injection attack against DeepID model in Figure 18. Our observations on Accessory Injection strategy are similar to the ones on Blended Accessory Injection strategy, which show that physically implementable backdoors can be successfully created with 20-40 poisoning samples injected into the training set. We also observe that the attack success rate is higher than using the Blended Accessory Injection strategy, and it requires fewer poisoning samples to achieve the same attack success rate.

The results of the Accessory Injection attacks against VGG-Face can be found in Figure 19. We observe similar phenomena using the Accessory Injection strategy as using the Blended Accessory Injection strategy discussed in Section VI. In particular, the effectiveness of physical attacks against VGG-Face model is similar to the DeepID model. For example, using the reading glasses pattern, after injecting 40 poisoning samples, Person 4 can create the backdoors with 80% attack success rate, which is even fewer than the ones needed to poison the DeepID model, for which even injecting 200 poisoning samples can only reach a 40% attack success rate. Meanwhile, the attack success rate using Accessory Injection strategy is higher than the Blended Accessory Injection strategy.

[2]http://www.robots.ox.ac.uk/∼vgg/software/vgg_face/

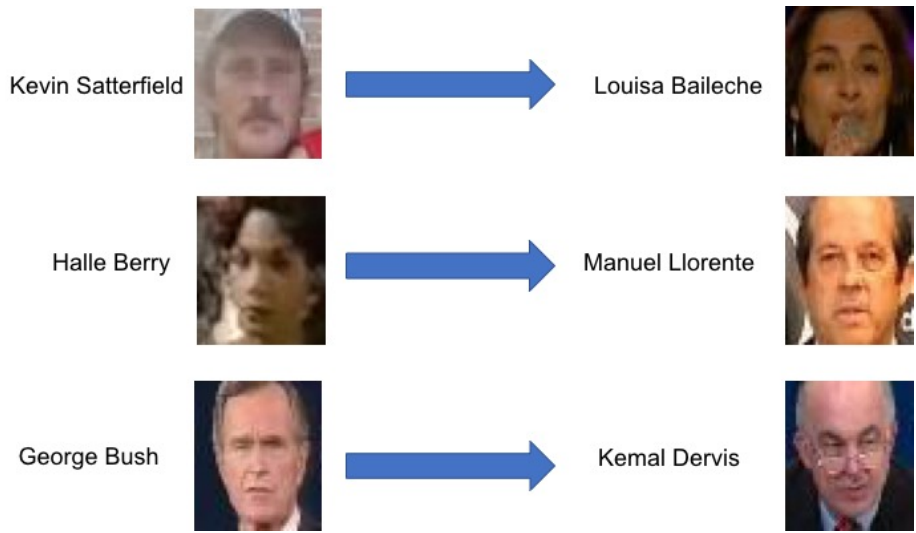

Fig. 17: Examples of input-instance-key attacks performed in our experiments, where for each image, its ground truth label is shown to its left. In this figure, face images on the left are the backdoor instances, and images on the right are those belonging to the target labels. We demonstrate that adding 5 poisoning samples into the training set is sufficient to mislead the model to predict the target label for the backdoor instances.

In a word, the attacker can perform successful physical attacks using the Accessory Injection strategy against both the DeepID model and the VGG-Face model, and the VGG-Face model pre-trained with auxiliary pristine data does not provide better resilience against backdoor attacks.

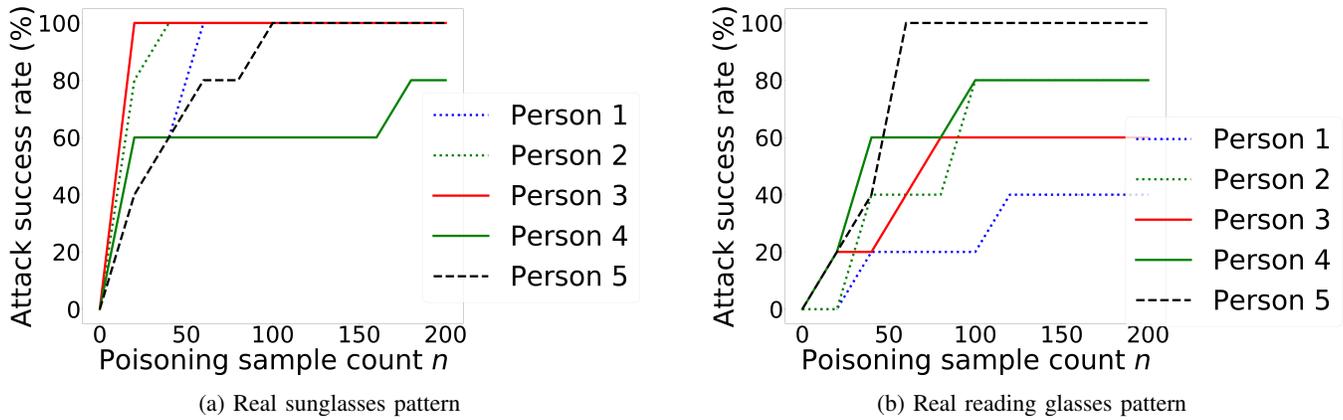

(a) Real sunglasses pattern

(b) Real reading glasses pattern

Fig. 18: Attack success rates of Accessory Injection attacks with physical patterns on DeepID model.

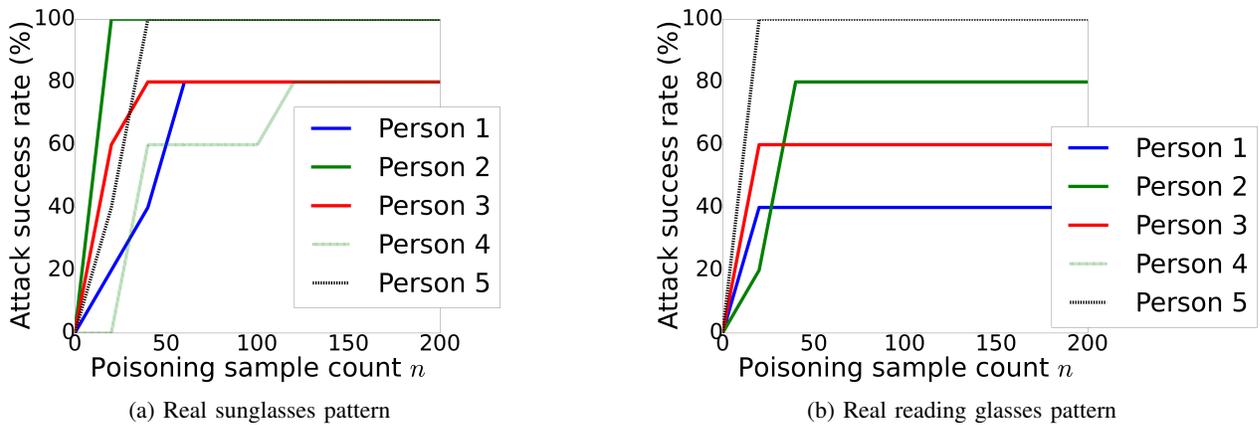

(a) Real sunglasses pattern

(b) Real reading glasses pattern

Fig. 19: Attack success rates of Accessory Injection attacks with physical patterns on VGG-Face model. In the right figure, lines of Person 2 and Person 4 overlap with each other.